\documentclass[conference]{IEEEtran}
\IEEEoverridecommandlockouts
\usepackage[normalem]{ulem}
\usepackage{cite}
\usepackage{amsmath,amssymb,amsfonts}
\usepackage{algorithmic}
\usepackage{graphicx}
\usepackage{textcomp}
\usepackage{xcolor}
\usepackage{listings}

\usepackage{booktabs}     
\usepackage{multirow}    
\usepackage{caption}     
\usepackage{graphicx}    
\usepackage{array}       
\usepackage{caption}
\usepackage{array}

\def\BibTeX{{\rm B\kern-.05em{\sc i\kern-.025em b}\kern-.08em
    T\kern-.1667em\lower.7ex\hbox{E}\kern-.125emX}}

\usepackage[absolute]{textpos}

\begin{document}

\begin{textblock}{5}(11.8,0.8)
(Special Session)
\end{textblock}


\title{TimelyHLS: LLM-Based Timing-Aware and Architecture-Specific FPGA HLS Optimization}

\author{\IEEEauthorblockN{Nowfel Mashnoor, Mohammad Akyash, Hadi Kamali, Kimia Azar}
\IEEEauthorblockA{\textit{Department of Electrical and Computer Engineering (ECE), University of Central Florida, Orlando, FL 32816, USA} \\
\{nowfel.mashnoor, mohammad.akyash, kamali, azar\}@ucf.edu}
}

\maketitle

\begin{abstract}

Achieving timing closure and design-specific optimizations in FPGA-targeted High-Level Synthesis (HLS) remains a significant challenge due to the complex interaction between architectural constraints, resource utilization, and the absence of automated support for platform-specific pragmas. In this work, we propose TimelyHLS, a novel framework integrating Large Language Models (LLMs) with Retrieval-Augmented Generation (RAG) to automatically generate and iteratively refine HLS code optimized for FPGA-specific timing and performance requirements. TimelyHLS is driven by a structured architectural knowledge base containing FPGA-specific features, synthesis directives, and pragma templates. Given a kernel, TimelyHLS generates HLS code annotated with both timing-critical and design-specific pragmas. The synthesized RTL is then evaluated using commercial toolchains, and simulation correctness is verified against reference outputs via custom testbenches. TimelyHLS iteratively incorporates synthesis logs and performance reports into the LLM engine for refinement in the presence of functional discrepancies. Experimental results across 10 FPGA architectures and diverse benchmarks show that TimelyHLS reduces the need for manual tuning by up to 70\%, while achieving up to 4× latency speedup (e.g., 3.85× for Matrix Multiplication, 3.7× for Bitonic Sort) and over 50\% area savings in certain cases (e.g., 57\% FF reduction in Viterbi). TimelyHLS consistently achieves timing closure and functional correctness across platforms, highlighting the effectiveness of LLM-driven, architecture-aware synthesis in automating FPGA design.


\end{abstract}

\section{Introduction}

FPGAs provide a flexible platform for accelerating diverse algorithms, but achieving timing closure on FPGAs remains a persistent challenge \cite{yanghua2016improving}. Timing closure entails ensuring that all signal paths meet the FPGA’s timing constraints (setup/hold times, clock delays, etc.), and it is critical for correct operation at the target clock frequency \cite{cong2022fpga}. In practice, reaching timing closure is an iterative and complex process, hindered by high clock rates, large and interconnected designs, and physical effects like routing delays \cite{numan2020towards}. Modern FPGA design flows often demand manual tuning and optimization to meet timing, especially when using High-Level Synthesis (HLS) tools to generate hardware from C/C++ code \cite{ustun2019lamda, sun2022correlated}.

HLS tools such as Xilinx Vitis HLS \cite{amd_vitis_hls} raise the design abstraction to high-level code, but they still rely heavily on user guidance to produce efficient, timing-compliant hardware \cite{xiong2024hlspilot}. In particular, performance-critical optimizations (e.g. loop pipelining, loop unrolling, memory partitioning) are typically controlled by pragmas or directives embedded in the HLS source \cite{pouget2025automatic}. Selecting the right combination of pragmas for a given design and FPGA is a non-trivial task that often requires deep hardware expertise \cite{chi2022democratizing}. 

Currently, there is a lack of automated support within HLS tools for platform-specific pragmas (i.e. directives tailored to a specific FPGA architecture or vendor). Each FPGA platform introduces unique resources and constraints, which often necessitates different pragmas or coding styles. Designers must manually adapt and tune their HLS code for each target device, since a pragma that works well (or is even recognized) on one toolchain may not apply on another \cite{lahti2025high}. For instance, a designer targeting Xilinx FPGAs might use the PIPELINE pragma to improve loop initiation intervals, while Intel’s HLS compiler requires a different pragma (ivdep) or coding convention to achieve similar results \cite{amd_vitis_hls_pipeline}. 

To ease the challenges of HLS optimization, researchers have developed automated methods like Design Space Exploration (DSE) frameworks \cite{liu2019accelerating, ferretti2018lattice}, which search large pragma spaces to find configurations with good Quality of Results (QoR). Tools such as AutoDSE \cite{sohrabizadeh2022autodse} use heuristic searches but require many HLS runs, making them slow. Analytical methods, like formulating pragma selection as a non-linear optimization problem \cite{pouget2025automatic}, reduce this cost by pruning poor choices. ML-based tools (e.g., HARP \cite{sohrabizadeh2023robust}) predict performance to guide optimizations more efficiently. However, models often struggle to generalize to new designs or architectures \cite{lahti2025high}. Despite progress, fully automated, widely adopted solutions remain lacking, especially for timing-driven optimization, leaving designers to rely on manual tuning.

Recent advances in Large Language Models (LLMs) have opened new possibilities in automating HLS optimization for FPGA design \cite{prakriya2025lift, xiong2024hlspilot, gautier2022sherlock}. Tools like LIFT \cite{prakriya2025lift} fine-tune LLMs with Graph Neural Network (GNN) analyses to insert pragmas, achieving up to 3.5× speedup over prior methods. HLSPilot \cite{xiong2024hlspilot} uses in-context learning and retrieval from vendor docs, combining DSE and profile-guided refinement to produce results comparable to expert designs. However, prompting general LLMs without domain-specific context can significantly degrade performance \cite{peng2023check}, demonstrating the need for integrated domain knowledge and feedback.

Despite recent progress, HLS-based FPGA design continues to face two critical challenges: (i) \textit{\uline{ensuring that synthesized designs meet strict timing requirements}}, which is complicated by factors like deep logic pipelines, routing congestion, and critical paths that are difficult to predict and optimize at a high level; and (ii) \textit{\uline{adapting optimizations to the unique characteristics of each FPGA architecture}}, where vendor-specific toolchains, resource constraints, and low-level features often necessitate custom pragma configurations and design patterns that do not generalize across platforms. These issues contribute to a persistent “closure gap” that current tools struggle to overcome without huge manual intervention.

To address these limitations, we propose TimelyHLS, a framework that combines LLMs with retrieval-augmented generation (RAG) and iterative refinement. TimelyHLS is guided by a structured knowledge base encoding FPGA-specific features, pragmas, and optimization heuristics. The LLM queries this knowledge during inference to generate HLS code tailored to the target architecture. This initial generation is followed by an iterative loop: synthesized RTL is evaluated using commercial tools, testbenches verify functional correctness, and synthesis logs are fed back into the model. The LLM then revises the code based on this feedback until the design achieves timing closure and correctness\footnote{While TimelyHLS mainly focuses on improving timing closure with functional correctness, the same flow can be used for power/area efficiency (if targeted and needed by the design specification).}. We evaluate TimelyHLS on various FPGA devices and benchmark kernels. Results show that TimelyHLS reduces manual iterations, consistently meets timing constraints, and delivers performance comparable to hand-tuned designs. Our contributions include: 

\noindent \textbf{(i) LLM + RAG for FPGA HLS:} We propose \textit{TimelyHLS}, the first framework to integrate an LLM with RAG for FPGA-specific HLS code generation. By grounding the model in a curated knowledge base of FPGA-specific architectural features, synthesis directives, and pragma strategies, TimelyHLS generates HLS code tailored to the target device.

\noindent \textbf{(ii) Iterative Refinement with Tool Feedback:} TimelyHLS employs an iterative refinement loop where synthesis reports, timing analysis, and functional verification feedback are reintegrated into the LLM. This enables the model to progressively resolve timing violations and improve design quality in a closed-loop, minimizing the need for manual tuning.

\noindent \textbf{(iii) Effective Timing Closure/Optimization:} Extensive experiments show that TimelyHLS achieves timing closure at an acceptable oveheard (area) across diverse FPGAs while significantly reducing manual intervention. It delivers performance and overall QoR (latency, area, timing) on par with, and in many cases exceeding, expert-optimized HLS designs.

\section{Related Works}


Early work on automating HLS optimization used heuristic search and static modeling to explore the design space of synthesis directives (pragmas) \cite{ansel2014opentuner}\footnote{Exhaustive search is infeasible as the combination of directives, e.g., loop unrolling, pipelining, memory partitioning, etc. grows exponentially, so studies applied heuristics to guide DSE without brute force.}. OpenTuner \cite{ansel2014opentuner}, a general auto-tuning framework adapted for HLS that orchestrates many search strategies (greedy, genetic algorithms, simulated annealing, etc.) via a multi-armed bandit approach. By dynamically choosing among strategies, OpenTuner effectively navigated large pragma spaces, often finding better solutions than any single algorithm alone. AutoDSE \cite{sohrabizadeh2022autodse}, a DSE tool that iteratively tunes one pragma at a time, always addressing the current performance bottleneck. By focusing on the most critical optimization step-by-step, AutoDSE achieved expert-level results with far fewer directives. Many DSE frameworks employed simulated annealing \cite{ding2024efficient} or hill-climbing \cite{pham2015exploiting} to automate pragma selection. These rule-based searches marked a big step in reducing engineering effort. However, purely heuristic approaches can miss global optima or get stuck in local optima, especially in very large parameter spaces.

Complementing above search techniques, analytical modeling approaches sought to speed up exploration by predicting HLS outcomes without full synthesis. Tools like Lin-Analyzer \cite{zhong2016lin}, COMBA \cite{zhao2017comba}, and more recently ScaleHLS \cite{ye2022scalehls}, use static code analysis (e.g., loop dependence graphs, pipeline initiation interval formulas) to estimate a design’s latency and resource usage under different pragma choices. By evaluating design points with these mathematical models, unpromising configurations can be pruned orders of magnitude faster than actually synthesizing them. In practice, these models often need re-tuning for each tool or hardware target, and they handle only a subset of possible pragmas or code structures (to keep the formulas tractable). As a result, analytical estimation is usually used as a component in a larger system.


Recent work has explored learning-based methods. ML frameworks like HARP \cite{sohrabizadeh2023robust} use surrogate models, often GNNs to predict performance from code and pragma configurations, enabling rapid DSE without full synthesis. Bayesian optimization (BO) tools like Sherlock \cite{gautier2022sherlock} further improve efficiency by prioritizing high-potential candidates through probabilistic modeling, excelling in multi-objective tuning.

Reinforcement learning (RL) frames HLS as sequential decision-making, where agents learn to apply transformations or insert pragmas to improve performance. Methods like AutoAnnotate \cite{shahzad2024autoannotate} report up to 4× speedup, though RL faces challenges with large design spaces and training times. Hybrid solutions like AutoHLS \cite{ahmed2023autohls} combine neural prediction with BO to prune poor candidates, achieving up to 70× speedup.

\begin{figure*}[t]
\centering
\includegraphics[width=\linewidth]{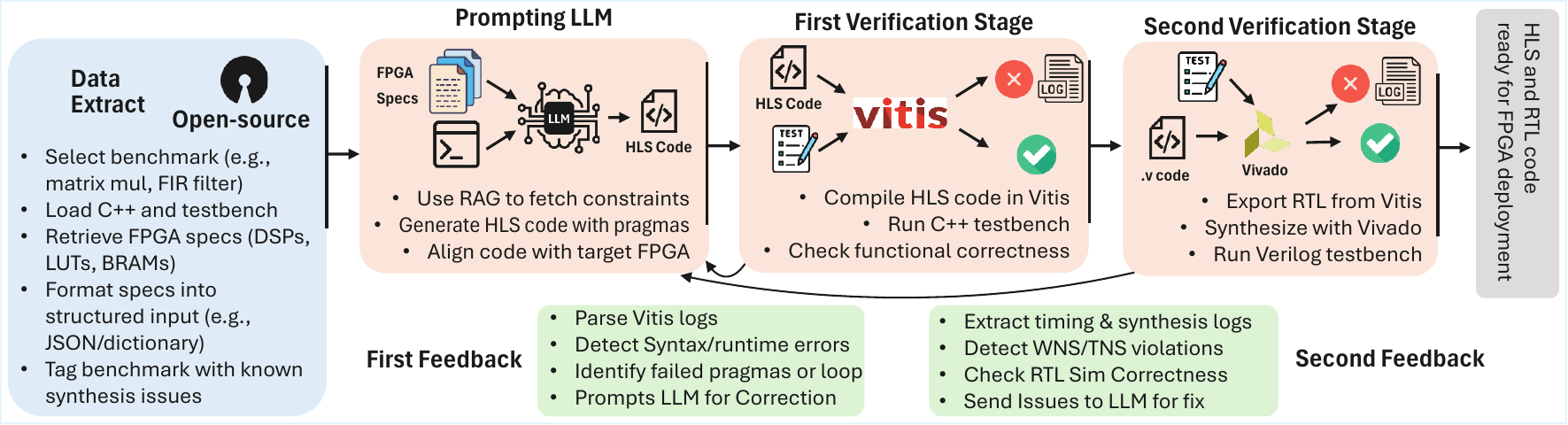}
\caption{Overview of TimelyHLS Framework.}
\vspace{-15pt}
\label{fig:timlyhls_framework}
\end{figure*}

LLM-based approaches, which are heavily in use in Register Transfer Language (RTL) design and verification \cite{akyash2024evolutionary, mashnoor2025llm, akyash2025rtl++}, aim to generate optimized HLS code directly. While general-purpose models like GPT-4 struggle without domain grounding, LIFT \cite{prakriya2025lift} improves results by fine-tuning on 40k+ annotated HLS samples with graph-based features, achieving 3.5× speedup over baselines. However, it lacks adaptability to new platforms and toolchains. HLSPilot \cite{xiong2024hlspilot} uses RAG, guiding the LLM with relevant HLS examples and vendor-specific rules at inference time. It enables structural code transformations and rivals expert performance. 

Building on this, proposed TimelyHLS enhances adaptability by dynamically querying an evolving, platform-aware knowledge base during generation. This ensures that all optimizations are context-aware, verifiable, and tailored to current toolchains, avoiding hallucinated or outdated directives.

\section{Overview of TimelyHLS}

The TimelyHLS framework automates timing-aware HLS code generation through a combination of LLM, RAG, and iterative synthesis feedback. Fig. \ref{fig:timlyhls_framework}, shows the top view of TimelyHLS framework. The workflow operates in two main verification stages, HLS-level and RTL-level, and leverages FPGA-specific knowledge to guide the model toward platform-compliant and timing-closure-friendly designs. To accomplish optimization for speed-up, the TimelyHLS framework consists of four main components, which are as follows: 

\subsection{Dataset Collection}

To establish (and evaluate) our framework, we curated a dataset of real-world HLS design examples gathered from open-source repositories, e.g., CHStone \cite{chstone}, LegUp benchmarks \cite{canis2013legup}, and MachSuite \cite{reagen2014machsuite}. These repositories provide diverse C/C++ programs widely used in HLS, covering a range of computational domains. We selected 10 representative HLS applications that reflect common optimization bottlenecks in FPGA synthesis, including timing violations, long critical paths, inefficient pipelining, and suboptimal resource allocation. A detailed description of each benchmark and its associated synthesis challenge is provided in Table~\ref{tab:hls_benchmarks}.

For each benchmark, we developed corresponding HLS C++ source files as well as custom testbenches to verify functional correctness during simulation and synthesis. All designs were compiled and analyzed using the Xilinx Vitis HLS toolchain. To ensure architectural diversity and practical relevance, we evaluated each design across 10 distinct FPGA targets spanning multiple device families. For each FPGA architecture, we collected synthesis reports, timing summaries, and resource utilization logs. Additionally, we captured the output of testbench simulations to verify functional correctness. This comprehensive dataset, which includes source code, testbenches, and tool-generated logs across multiple architectures, forms the basis for evaluating the effectiveness of our proposed LLM-based (prompting) TimelyHLS framework.

\subsection{Prompting LLM for Initial Code Generation}

\begin{table}[t]
\scriptsize
\centering
\caption{HLS Selected Benchmarks with Synthesis Challenges (Timing-wise) Targeted and Used in TimelyHLS.}
\label{tab:hls_benchmarks}
\setlength\tabcolsep{2pt}
\begin{tabular}{@{} p{85pt} p{160pt} @{}}
\toprule  
Application & Optimization Challenge \\
\cmidrule(r){1-1}\cmidrule(r){2-2}
Matrix Multiplication & Long critical path due to nested loops; loop pipelining inefficiencies. \\
\cmidrule(r){1-1}\cmidrule(r){2-2}
Convolution & Timing violations caused by inefficient memory access and computation overlap. \\
\cmidrule(r){1-1}\cmidrule(r){2-2}
Vector Dot Product & Underutilized resources and insufficient parallelism. \\
\cmidrule(r){1-1}\cmidrule(r){2-2}
Vector Addition & Suboptimal loop unrolling with moderate timing slack. \\
\cmidrule(r){1-1}\cmidrule(r){2-2}
Bitonic Sort & Deep logic pipelines leading to routing congestion and critical path delays. \\
\cmidrule(r){1-1}\cmidrule(r){2-2}
Viterbi Decoder & Control dependencies causing resource contention. \\
\cmidrule(r){1-1}\cmidrule(r){2-2}
Adaptive FIR Filter (LMS) & Feedback loop latency and failed timing closure due to iteration dependencies. \\
\cmidrule(r){1-1}\cmidrule(r){2-2}
CORDIC Algorithm & Inefficient pipelining due to iterative data dependencies. \\
\cmidrule(r){1-1}\cmidrule(r){2-2}
Matrix-Vector Multiplication & Memory partitioning bottlenecks leading to timing degradation. \\
\cmidrule(r){1-1}\cmidrule(r){2-2}
Needleman–Wunsch (DP) & Irregular memory access patterns causing critical path delay and low throughput. \\
\bottomrule
\end{tabular}
\end{table}

For each sample in our dataset, we craft a task-specific prompt that describes the functionality and performance objectives of the design (e.g., loop behavior, target throughput, or memory access constraints). This prompt is paired with target FPGA metadata (i.e. device family, number of DSPs, BRAMs, LUTs, and timing constraints as part of a RAG pipeline). The architectural specifications are embedded from datasheets and vendor tool documentation into a structured knowledge base.

We then query the LLM (e.g., Code LLaMA or GPT-4) with the prompt and retrieved FPGA constraints to generate HLS-compliant C/C++ code. The model is expected to insert relevant synthesis directives (pragmas) such as \texttt{\#pragma HLS pipeline}, \texttt{unroll}, or \texttt{array\_partition}, aligned with the resource capabilities of the target device.

\subsection{HLS-Level Verification and Correction}

The generated HLS code is compiled and simulated using Xilinx Vitis HLS\footnote{All prompts and scripts are configurable (parametrized) to be reusable for different vendors (to be easily used with different toolsets.}, paired with the testbench we previously crafted for each benchmark. This first stage ensures that the model's output is functionally correct and synthesizable at the C level. If the compilation fails or the functional simulation does not produce expected results, we extract relevant information from Vitis logs (e.g., syntax errors, resource binding issues, or pipeline depth violations) and return this feedback to the LLM in the form of an augmented prompt. The LLM is asked to revise its output to address the specific failures. This process is repeated \textbf{iteratively} until the design passes both HLS synthesis and functional simulation.

\subsection{RTL-Level Verification and Timing Evaluation}

Once the HLS design passes the first verification stage, we export the RTL (Verilog) output and proceed with the second stage using Xilinx Vivado. In this phase, we generate the corresponding Verilog testbenches and synthesize the design for the selected FPGA target. Vivado's post-synthesis reports are then used to evaluate timing closure (e.g., Worst Negative Slack (WNS), Total Negative Slack (TNS)), resource utilization, and syntactic validity of the RTL. We also simulate the design at the RTL level using the generated testbenches to verify behavioral equivalence with the HLS-level outputs. If the synthesizer (i.e., Xilinx Vivado) fails to synthesize the design or simulation results deviate from the expected output, we extract detailed logs, including synthesis errors, critical path reports, and functional mismatches and pass them as feedback to the LLM. This closes the second loop of iterative refinement, allowing the model to correct deeper architectural or low-level issues not observable during HLS.

This two-stage refinement loop continues until the design satisfies these criteria: (i) Passes functional simulation in both HLS and RTL levels; (ii) Is synthesizable by Vivado for the target FPGA architecture; and (iii) Meets timing closure requirements with no negative slack.

By integrating FPGA-specific architectural guidance into generation and leveraging compiler logs as feedback, TimelyHLS transforms the traditional trial-and-error-based HLS optimization into an automated, LLM-driven pipeline.

\section{Experimental Setup}
\subsection{Experimental Environment and Tools}
All experiments were conducted on a Linux-based development environment (Ubuntu 24.04.2 LTS) using Xilinx Vitis HLS (Version 2024.2) for HLS and Vivado (Version 2024.2) for RTL synthesis and implementation. The experimental infrastructure consisted of 13th Gen Intel(R) Core(TM) i7-13700 Processor and 32GB of memory capacity to accommodate the synthesis flows, parallel DSE, and iterative LLM inference processes. We evaluated TimelyHLS across 10 diverse FPGA devices, including Artix-7, Spartan-7, Zynq, and Virtex UltraScale+, to ensure comprehensive architectural coverage. The selected devices represent a wide spectrum of resource capacities, from low-cost embedded solutions to high-end  accelerators listed in Table \ref{tab:fpga_families}. All designs were synthesized with constraints of achieving the maximum frequency.

\begin{table}[t]
\scriptsize
\centering
\caption{Targeted FPGA Families and their Applications.}
\label{tab:fpga_families}
\setlength\tabcolsep{2pt}
\begin{tabular}{@{} p{60pt} p{90pt} p{90pt} @{}}
\toprule  
FPGA Family & Part Number(s) & Typical Applications \\
\cmidrule(r){1-1}\cmidrule(r){2-2}\cmidrule(r){3-3}
Zynq & xc7z020-clg484-1 & Embedded applications \\
\cmidrule(r){1-1}\cmidrule(r){2-2}\cmidrule(r){3-3}
Zynq UltraScale+ & xczu3eg-sbva484-1-e & Heterogeneous computing \\
\cmidrule(r){1-1}\cmidrule(r){2-2}\cmidrule(r){3-3}
Artix/Kintex-7 & xc7a200tfbg676-2, & Cost-optimized designs \\
& xc7k325tffg676-2 & \\

\cmidrule(r){1-1}\cmidrule(r){2-2}\cmidrule(r){3-3}
Spartan-7 & xc7s50-ftgb196-2 & Ultra-low-cost applications \\
\cmidrule(r){1-1}\cmidrule(r){2-2}\cmidrule(r){3-3}
Virtex UltraScale+ & xcvu9p-flgb2104-2-e, & High-performance computing \\
& xcvu11p-flga2577-1-e, \\
& xcvu9p-flgb2104-1-e \\
\cmidrule(r){1-1}\cmidrule(r){2-2}\cmidrule(r){3-3}
Kintex UltraScale+ & xck26-sfvc784-2LV-c & Balanced performance-power \\
\cmidrule(r){1-1}\cmidrule(r){2-2}\cmidrule(r){3-3}
Versal AI Edge & xave2602-nsvh1369-1LJ-i-L & AI/ML acceleration \\
\bottomrule
\end{tabular}
\end{table}

\subsection{Large Language Model Configuration}

We conducted comparative experiments using two state-of-the-art LLMs: OpenAI GPT-4 and Anthropic Claude-3.5-Sonnet, both accessed via their respective APIs with default temperature settings (0.7) to balance creativity and determinism in code generation. The RAG knowledge base was constructed by extracting and structuring information from official FPGA datasheets, vendor HLS user guides, and architectural reference manuals for each target device family.

\section{Results and Evaluation}

To evaluate TimelyHLS, the benchmarks span a variety of domains, e.g., linear algebra, signal processing, sorting, and dynamic programming, each presenting unique challenges such as long critical paths, deep loop dependencies, or inefficient memory access. The evaluation focuses on four key metric categories: timing and performance, resource utilization, loop-level optimizations, and structural design changes.

\subsection{Performance Improvements}

\begin{figure}[b]
\centering
\includegraphics[width=\linewidth]{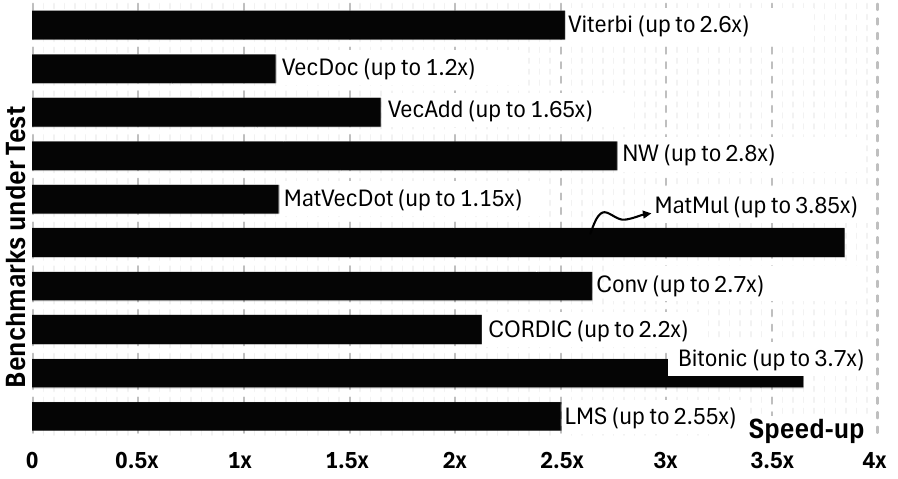}
\caption{Performance Speedup Ratios for Artix 7.}
\label{fig:speed_up_chart}
\end{figure}

TimelyHLS demonstrated significant performance gains across several benchmarks. As illustrated in Fig. \ref{fig:speed_up_chart}, speedup values on Artix-7 reached up to 4×, with applications like Matrix Multiplication, LMS Filter, and Bitonic Sort showing the most notable improvements. These gains are primarily the result of architecture-specific pipelining strategies and effective insertion of pragmas such as pragma HLS pipeline and loop unrolling. These changes reduce the initiation interval (II) and enable higher parallelism.

\subsection{Timing and Latency Analysis}
The timing closure performance of TimelyHLS is substantiated by latency and slack results presented in Table \ref{tab:latency_std}. Across multiple benchmarks, the framework not only resolved negative slack violations but also preserved optimal latency characteristics when further improvement was infeasible or unnecessary. As shown, in the Matrix Multiplication benchmark, the baseline design exhibited negative slack, a direct result of deep loop nesting and non-optimized memory access, which introduced excessive combinational delay. TimelyHLS resolved this by restructuring loop hierarchies and selectively applying pipelining with loop unrolling and array partitioning, resulting in a fully timing-closed implementation with zero slack. Similarly. for the Vector Dot Product, the baseline implementation underutilized available DSP resources and suffered from inefficient loop scheduling, leading to intermittent timing violations. TimelyHLS addressed this by rebalancing the loop-carried dependencies and improving the data flow via pipelining and unrolling. 

\begin{table}[t]
\scriptsize
\centering
\setlength\tabcolsep{2.5pt}
\caption{Latency Comparison: Base vs. TimelyHLS.}
\label{tab:latency_std}
\begin{tabular}{@{} l *{9}c @{}}
\toprule
 & Bitonic & CORDIC & Mat-Vec & Mat Mul & VecAdd & VecDot & Viterbi \\
\cmidrule(r){1-1}\cmidrule(r){2-2}\cmidrule(r){3-3}\cmidrule(r){4-4}\cmidrule(r){5-5}\cmidrule(r){6-6}\cmidrule(r){7-7}\cmidrule(r){8-8}
\textbf{Base (ns)} &  0.1 & 2.7 & 0.54 & -0.08 & 2.7 & -0.54 & 2.7 \\
\textbf{TimelyHLS (ns)} &  0.1 & 2.7 & 0.62 & 0.1 & 2.7 & 0.54 & 2.7 \\
\bottomrule
\end{tabular}
\end{table}

\subsection{Resource Utilization and Performance Tradeoffs}

\begin{table}[b]
\scriptsize
\centering
\caption{Flip-Flop (FF) Usage  across FPGA Families.}
\label{tab:ff_usage_comparison}
\setlength\tabcolsep{9pt}
\begin{tabular}{@{}llrrr@{}}
\toprule
\textbf{Benchmark} & \textbf{Family} & \textbf{Part} & \textbf{FF Used} & \textbf{FF Change (\%)} \\
\cmidrule(r){1-1}\cmidrule(r){2-2}\cmidrule(r){3-3}\cmidrule(r){4-4}\cmidrule(r){5-5}
Viterbi & Artix-7 & xc7a200t & 247 & -57.34 \\
Viterbi & Spartan-7 & xc7s50 & 247 & -57.34 \\
CORDIC & Spartan-7 & xc7s50 & 329 & -12.27 \\
Vec Dot & Artix-7 & xc7a200t & 572 & 22.75 \\
Vec Add & Zynq & xc7z020 & 2479 & 38.11 \\
Vec Add & Spartan-7 & xc7s50 & 2468 & 39.91 \\
Vec Add & Virtex-U+ & xcvu11p & 2479 & 38.11 \\
\bottomrule
\end{tabular}
\end{table}

\begin{table}[b]
\scriptsize
\centering
\caption{LUTs Usage across FPGA Families.}
\label{tab:lut_usage_comparison}
\setlength\tabcolsep{6pt}
\begin{tabular}{@{}llrrr@{}}
\toprule  
\textbf{Benchmark} & \textbf{Family} & \textbf{Part} & \textbf{LUTs Used} & \textbf{LUTs Change (\%)} \\
\cmidrule(r){1-1}\cmidrule(r){2-2}\cmidrule(r){3-3}\cmidrule(r){4-4}\cmidrule(r){5-5}
Viterbi Dec. & Artix-7 & xc7a200t & 619 & -48.24 \\
Viterbi Dec. & Spartan-7 & xc7s50 & 647 & -47.91 \\
Vec. Dot Prod. & Artix-7 & xc7a200t & 669 & 40.84 \\
Vec. Addition & Spartan-7 & xc7s50 & 2460 & 46.43 \\
Viterbi & Versal AI & xcvc2602 & 479 & 0.00 \\
Vec. Addition & Virtex-U+ & xcvu11p & 2578 & 52.82 \\
Viterbi Dec. & Virtex-U+ & xcvu11p & 1946 & 58.47 \\
Mat-Vec Mult. & Artix-7 & xc7a200t & 3132 & 65.18 \\
Mat-Vec Mult. & Spartan-7 & xc7s50 & 3132 & 65.18 \\
\bottomrule
\end{tabular}
\end{table}

Tables \ref{tab:ff_usage_comparison} and \ref{tab:lut_usage_comparison} reflect the impact of TimelyHLS on balancing hardware area consumption, measured through flip-flop (FF) and lookup table (LUT) usage, with improvements in latency and timing closure. The observed trade-offs underscore the framework's architecture-aware optimization strategy, where moderate increases in resource utilization achieve the latency improvements, while in other cases, aggressive area savings are prioritized when latency is already near-optimal.

In benchmarks such as vector addition and matrix-vector multiplication, TimelyHLS used deeper pipelining and memory partitioning to eliminate bottlenecks and sustain loop throughput. While this increased LUT usage by 50–65\%, improvements in timing slack and loop initiation intervals justified the added area (deliberate trade-off: reducing latency with parallelism and logic duplication increases area). On the other hand, for benchmarks like the Viterbi Decoder, TimelyHLS reduced FF and LUT usage by over 50\%. This implies that the original design had redundant control logic or non-optimized datapath that could be compacted without affecting latency. These observations highlight the context-sensitive nature of the latency–area trade-off. TimelyHLS adapts its optimization to the designs' needs, aggressively optimizing for performance when needed, and concentrating on compactness when further gains are unnecessary. All final designs met FPGA resource constraints, showing both adaptability and architectural feasibility.

\subsection{Loop-Level Optimization Strategies}

\begin{figure}[b]
\centering
\includegraphics[width=\linewidth]{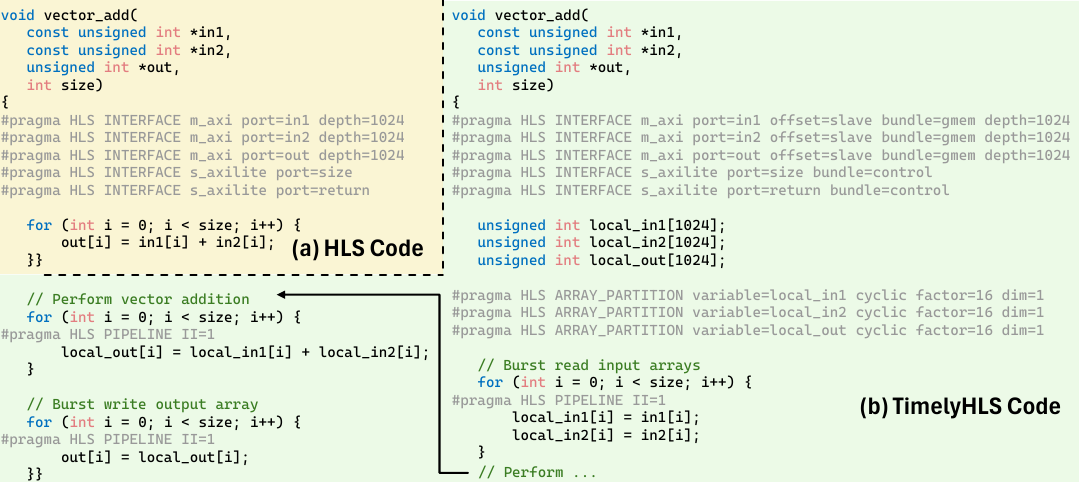}
\caption{TimelyHLS Code Example with Optimization.}
\label{fig:code_example}
\end{figure}

Table \ref{tab:timelyhls_loopii} shows the performance impact of TimelyHLS on key look initiation interval (II). As shown, in Matrix Multiplication, the initiation interval (II) was reduced from 16 to 1–2, significantly enhancing throughput. Bitonic Sort, previously limited by non-pipelined loops, was fully pipelined with II=1, resolving the primary performance bottleneck. Similarly, Matrix-Vector Multiplication and CORDIC benefited from loop unrolling and pipelining, leading to reduced latency and improved data movement. These results indicate that TimelyHLS effectively identifies loop-carried dependencies and applies appropriate directives to maximize hardware utilization and scheduling efficiency. An example of such impact has shown in Fig. \ref{fig:code_example} that represents code snippet of Vector Add (base vs. TimelyHLS implementation). 

\begin{table}[t]
\fontsize{7pt}{8pt}\selectfont
\centering
\caption{Impact of TimelyHLS on Loop II.}
\label{tab:timelyhls_loopii}
\setlength\tabcolsep{3pt}
\begin{tabular}{@{}p{65pt}p{30pt}p{35pt}p{85pt}@{}}
\toprule
\textbf{Project} & \textbf{Base} & \textbf{TimelyHLS} & \textbf{Reason} \\
\midrule
Matrix Multiplication & 16 & 1--2 & Faster throughput. \\
Bitonic Sort & Non-pip. & All II=1 & Loop pipelined. \\
Matrix-Vector Mult. & Not pip. & Pip. (II=1) & Added pipelining. \\
Vector Dot Product & 1 & 1 & Fixed timing. \\
Vector Addition & 1--2 & 1 & Higher BRAM usage. \\
CORDIC & 2 & Unrolled & Loop unrolled. \\
\bottomrule
\end{tabular}
\end{table}

\subsection{Structural Optimization across FPGA Architectures}

Evaluation across FPGA families (e.g., Artix-7) shows that TimelyHLS often restructures modules and interfaces to enhance performance. For instance, in Matrix Multiplication, latency dropped significantly, from 16,531 to 4,277 cycles, accompanied by the use of AXI interfaces for improved modularity. Similarly, Bitonic Sort saw a 3.6× speedup, enabled by the addition of sparsemux units for faster scheduling. In contrast, Vector Dot Product saw a modest latency increase (519 to 640 cycles) but a dramatic rise in DSP usage (5 to 160), indicating operator duplication to meet timing. Overall, structural changes via TimelyHLS often led to performance gains, though at the cost of complexity or resource overhead.

\subsection{Architectural Adaptability and Design Quality}

\begin{figure}[t]
\centering
\includegraphics[width=\linewidth]{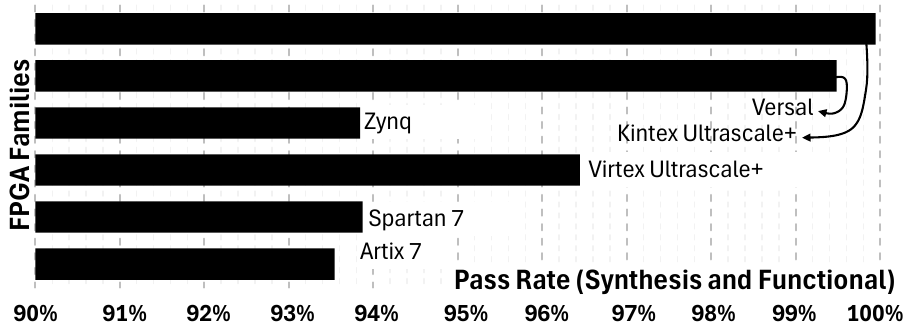}
\caption{Code Generation Success across FPGA Families.}
\label{fig:code_success_percentage}
\end{figure}

TimelyHLS demonstrates strong architectural adaptability, consistently generating synthesizable HLS code across a wide range of FPGA families, from low-cost devices like Spartan-7 and Artix-7 to high-end platforms such as Zynq UltraScale+ and Virtex UltraScale+ (see Fig. \ref{fig:code_success_percentage}). Most benchmarks compiled successfully across nearly all devices, indicating broad generalizability without the need for manual retargeting. Failures were mostly limited to complex designs with irregular memory access or feedback-heavy loops, which struggled on resource-constrained FPGAs. In contrast, high-end devices like the Virtex UltraScale+ consistently supported even the most challenging designs. This suggests that TimelyHLS adapts its code generation to match platform capabilities—favoring compact, efficient structures on smaller FPGAs and leveraging advanced features (e.g., AXI interfacing, etc.) on larger ones. 

\section{Conclusion and Future Work}

This paper presented TimelyHLS, a framework that combines large language models, retrieval-augmented generation, and synthesis feedback to automate timing-aware, architecture-specific HLS code generation for FPGAs. By leveraging a structured knowledge base, TimelyHLS produces functionally correct, synthesizable designs that meet timing constraints across a broad range of FPGA architectures. Experiments show that TimelyHLS generalizes well to both low-end and high-performance platforms, adapting its optimization strategies to balance latency, area, and throughput. It automates complex design transformation, while maintaining high synthesis success rates even on resource-limited devices. Future work will extend the framework to support multi-objective optimization (e.g., power, area, performance trade-offs), integrate additional toolchains and hardware platforms, and improve model generalization through fine-tuning and curriculum learning. 

\bibliographystyle{IEEEtran}
\bibliography{refs}

\end{document}